\documentclass[aps,prb,groupedaddress,
10pt
]{revtex4} 
\usepackage{graphicx,graphics,amsfonts,amssymb,color,bm,verbatim,ifsym,amsmath}

\usepackage[T1]{fontenc}       

\def\be{\begin{equation}}  
\def\ee{\end{equation}}  
\def\ba{\begin{eqnarray}}  
\def\ea{\end{eqnarray}}  

\begin{document} 

\title[Influence of optical nonlinearities on plasma waves in graphene]
  {Influence of optical nonlinearities on plasma waves in graphene}
\author{Sergey A. Mikhailov}
\affiliation{Institute of Physics, University of Augsburg, D-86135 Augsburg, Germany}
\email{sergey.mikhailov@physik.uni-augsburg.de}



\begin{abstract}
A theory of the nonlinear plasma waves in graphene is developed in the nonperturbative regime. The influence of strong electric fields on the position and linewidth of plasma resonances in the far-infrared 
transmission experiments, as well as on the wavelength and the propagation length in the scanning near-field optical microscopy  
experiments is studied. The theory shows that the fields of order of a few to a few tens of kV/cm should lead to a red shift and broadening of plasma resonances in the first type and to a reduction of the wavelength and the propagation length in the second type of experiments.
\end{abstract}

\maketitle

\section{Introduction}
The field of graphene plasmonics attracted much interest in recent years \cite{Low14}. Theoretically the spectrum of two-dimensional (2D) plasmons in doped graphene was calculated in Refs. \cite{Wunsch06,Hwang07}. In the long-wavelength limit $q\ll k_F$ the 2D plasmon frequency $\omega$ is related to the 2D plasmon wavevector $q$, electron density $n_s$, Fermi energy $E_F$ and the effective scattering time $\tau$ by the dispersion relation 
\be 
\omega\left(\omega+\frac 1\tau\right)= 
\frac{2e^2E_F}{\hbar^2\kappa_0}q = \frac{2e^2\sqrt{\pi n_s}v_Fq}{\hbar\kappa_0},\label{de}
\ee
where $k_F$ is the Fermi wavevector, $v_F\approx 10^8$ cm/s is the Fermi velocity and $\kappa_0$ is the dielectric constant of the surrounding medium. Experimentally, the 2D plasmons in graphene have been observed using the far-infrared (FIR) transmission spectroscopy in a system of narrow graphene stripes \cite{Ju11} and using the scanning near-field optical microscopy (SNOM) in Refs. \cite{Koppens11,Chen12,Fei12}; other experimental techniques have also been used, see Ref. \cite{Liu08}. In the first method \cite{Ju11} the plasmon wavevector $q$ in (\ref{de}) is given by the stripe width $W$, $q\simeq\pi/W$, and one observes a transmission resonance with the position $\omega'_p$ and the linewidth $\omega''_p$ determined by the real and imaginary parts of the frequency $\omega$ calculated from Eq. (\ref{de}). In the second technique \cite{Koppens11,Chen12,Fei12} the 2D plasmon frequency is determined by the frequency of the incident radiation and the wavelength $\lambda_p=2\pi/q'$ and the propagation length $L_p=1/q''$ of the plasmon are given by the real and imaginary parts of $q$ calculated from the dispersion equation (\ref{de}). 

Another topic actively developing in graphene optics nowadays is the nonlinear electrodynamic response of graphene. It was predicted in 2007 (Ref. \cite{Mikhailov07e}) that, due to the linear energy dispersion of graphene quasi-particles this material should demonstrate a strongly nonlinear electrodynamic response. Shortly after that it was confirmed by both experimental and further theoretical studies that the nonlinear parameters of graphene are much larger than in many other materials indeed (the strong nonlinearity of graphene was disputed in Ref. \cite{Khurgin14}; a detailed analysis of this paper can be found in a recent work \cite{Mikhailov17preprint}). Experimentally the higher harmonics generation \cite{Dragoman10,Bykov12,Kumar13,Hong13}, four-wave mixing\cite{Hendry10,Gu12}, Kerr effect \cite{Zhang12,Vermeulen16,Dremetsika16} have been measured by different methods. Theoretically the nonlinear electrodynamic response was studied both within the quasiclassical \cite{Mikhailov08a,Mikhailov09a} and quantum approaches\cite{Cheng15,Cheng16,Mikhailov16a,Semnani16}, and different aspects of the nonlinear graphene response have been analyzed \cite{Mikhailov09b,Mikhailov11c,Yao14,Smirnova14,MariniAbajo16,Savostianova17a}.

In view of the great interest to the two topics outlined above a question arises how the nonlinear properties of graphene influence the plasma waves in this material. The opposite question -- how plasmons influence the strength of the nonlinear effects in graphene -- has been discussed in Refs. \cite{Mikhailov11c,CoxAbajo14,CoxAbajo15}. Here we address the question how the frequency $\omega'_p$ and the linewidth $\omega''_p$ of graphene plasmons in the first type of experiments and the wavelength $\lambda_p$ and the propagation length $L_p$ in the second type of experiments are modified if the intensity of the plasmon electric field is so strong that the nonlinear effects become essential. 

\section{Theory\label{sec:theor}} 

In the linear electrodynamics the spectrum (\ref{de}) of plasma waves in 2D electron systems is calculated from the dispersion relation
\be 
\epsilon_{\rm lin}(q,\omega)\equiv 1+\frac{2\pi i \sigma(\omega) q}{\omega\kappa_0
}=0\label{defull}
\ee
where $\sigma(\omega)$ is the linear conductivity of the 2D layer and $\epsilon_{\rm lin}(q,\omega)$ is the effective dielectric function of graphene in the linear approximation. If graphene is doped and the plasmon frequency satisfies the condition $\hbar\omega\lesssim E_F$ (equivalent to $q\lesssim k_F$) the inter-band contribution to the linear conductivity of graphene (see Ref. \cite{Mikhailov07d}) can be neglected and $\sigma(\omega)$ in (\ref{defull}) is given by the intra-band conductivity
\be 
\sigma(\omega)\simeq \sigma^{\rm intra}(\omega) =\frac{e^2E_F}{\pi\hbar^2}\frac{i}{\omega+i/\tau}=
\frac{\sigma_0}{1-i\omega\tau},
\label{Drude}
\ee 
where $\sigma_0$ is the static conductivity of graphene. The Drude formula (\ref{Drude}) together with the dispersion equation (\ref{defull}) give the 2D plasmon spectrum (\ref{de}). 

The linear response approach is valid when the plasmon field is not very strong, i.e. when the field parameter
\be 
{\cal F}_\omega=\frac{eE_0}{\hbar k_F\omega} \label{Fomega}
\ee
is small as compared to unity, ${\cal F}_\omega\ll 1$, see Refs. \cite{Mikhailov07e,Mikhailov17a}. A nonperturbative quasiclassical theory which gives a general relation between the current and the field at arbitrary values of ${\cal F}_\omega$ has been recently developed in Ref. \cite{Mikhailov17a}. It was shown there that in the nonlinear regime the linear conductivity $\sigma(\omega)$ should be replaced by the function
\be 
\sigma_{\omega,\omega}(\omega\tau,{\cal F}_\tau)=
\sigma_0 {\cal S}_{1}(\omega\tau,{\cal F}_\tau),
\label{sigma(1)ww}
\ee
where 
\be 
{\cal F}_\tau=\omega\tau{\cal F}_\omega=\frac{eE_0\tau}{\hbar k_F}
\ee
is a frequency independent field parameter which is convenient to use analyzing the frequency dependencies of the nonlinear response,
\ba 
{\cal S}_{1}(\omega\tau,{\cal F}_\tau)=
\int^{\infty}_0 e^{-\xi} \frac{\sin(\omega\tau \xi/2)}{\omega\tau/2} 
B_{1}\left({\cal F}_\tau\frac{\sin(\omega\tau \xi/2)}{\omega\tau/2}\right)
e^{i\omega\tau \xi/2} d \xi
\label{S(1)}
\ea
is a complex function of $\omega\tau$ and ${\cal F}_\tau$,  
\ba 
B_{1}(a)=
\frac 4\pi\int_{0}^{\pi/2} \frac{\sin^2 x}{\sqrt{1+(a\sin x)^2} }
 \ 
_2F_1\left(\frac 14,\frac 34,2;\left(\frac{2 a\sin x}{1+(a\sin x)^2}\right)^2\right)  dx \label{B1}
\ea
and $_2F_1\left(a,b,c;x\right)$ is the hypergeometric function (for details see Ref. \cite{Mikhailov17a}). The dispersion equation of 2D plasmons in graphene in the nonlinear regime then reads
\be 
\epsilon_{\rm nonlin}(q,\omega)\equiv 1+\frac{2\pi i }{\omega\kappa_0}\frac{e^2}{\pi\hbar} \frac{E_F\tau}{\hbar} {\cal S}_{1}(\omega\tau,{\cal F}_\tau) q=0.
\ee
Now we can analyze the obtained results.

\section{Results and discussion\label{sec:result}}

\subsection{Nonlinearity in a FIR transmission experiment}

\begin{figure}
\includegraphics[width=0.5\textwidth]{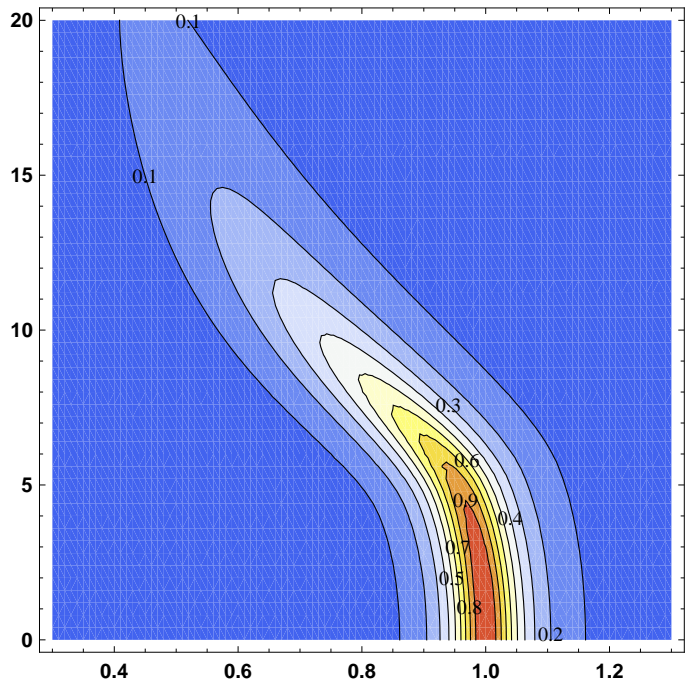}
\caption{\label{fig:abs} The nonlinear 2D plasmon absorption as a function of frequency $\omega/\omega_p$ (horizontal axis) and electric field parameter ${\cal F}_\tau$ (vertical axis) at $\omega_p\tau=10$.}
\end{figure}

In the FIR transmission (absorption) experiment, in the linear regime, the absorption coefficient is proportional to $A_{\rm lin}(\omega)\propto \sigma'(\omega)/|\epsilon_{\rm lin}(q,\omega)|^2$ where $q\simeq \pi/W$ is fixed. In the nonlinear regime we get $A_{\rm nonlin}(\omega)\propto \sigma'_{\omega,\omega}(\omega\tau,{\cal F}_\tau)/|\epsilon_{\rm nonlin}(q,\omega)|^2$ which gives
\be 
A_{\rm nonlin}(\omega)\propto \frac{ {\cal S}'_{1}(\omega\tau,{\cal F}_\tau)}
{\left|1+
i(\omega_p \tau)^2
\frac{{\cal S}_{1}(\omega\tau,{\cal F}_\tau)}{\omega\tau}\right|^2},
\label{absorption}
\ee
where
\be 
\omega_p^2=\frac{2 e^2 E_F}{\hbar^2\kappa_0}q
\label{plasm}
\ee
is the (linear-regime) plasma frequency, see Eq. (\ref{de}). Figure \ref{fig:abs} shows the nonlinear absorption coefficient (\ref{absorption}) as a function of frequency $\omega/\omega_p$ and the field parameter ${\cal F}_\tau$ at $\omega_p\tau= 10$. If ${\cal F}_\tau\to 0$ the absorption spectrum has a standard Drude shape with the quality factor of order of $10$. When the field parameter ${\cal F}_\tau$ grows but remains smaller than $\simeq\omega_p\tau$ the influence of the nonlinear effects is not essential: the resonance frequency experiences a red shift and the resonance becomes broader but these changes are not large. If ${\cal F}_\tau$ approaches the value $\omega_p\tau=10$ and exceeds it, the resonance frequency decreases dramatically, its linewidth grows and becomes comparable with the frequency. The boundary between the linear and nonlinear regimes is thus determined by the condition ${\cal F}_p\simeq 1$, where 
\be 
{\cal F}_p=\frac{{\cal F}_\tau}{\omega_p\tau}=\frac{eE_0}{\hbar k_F\omega_p}=\frac{eE_0v_F/\omega_p}{E_F}\label{Fp}
\ee
(compare with (\ref{Fomega})). The parameter ${\cal F}_p$ determines how much energy electrons obtain from the external electric field during one period of plasma oscillations as compared to their average (Fermi) energy. The nonlinear regime is realized when ${\cal F}_p\gtrsim 1$. 

\begin{figure}
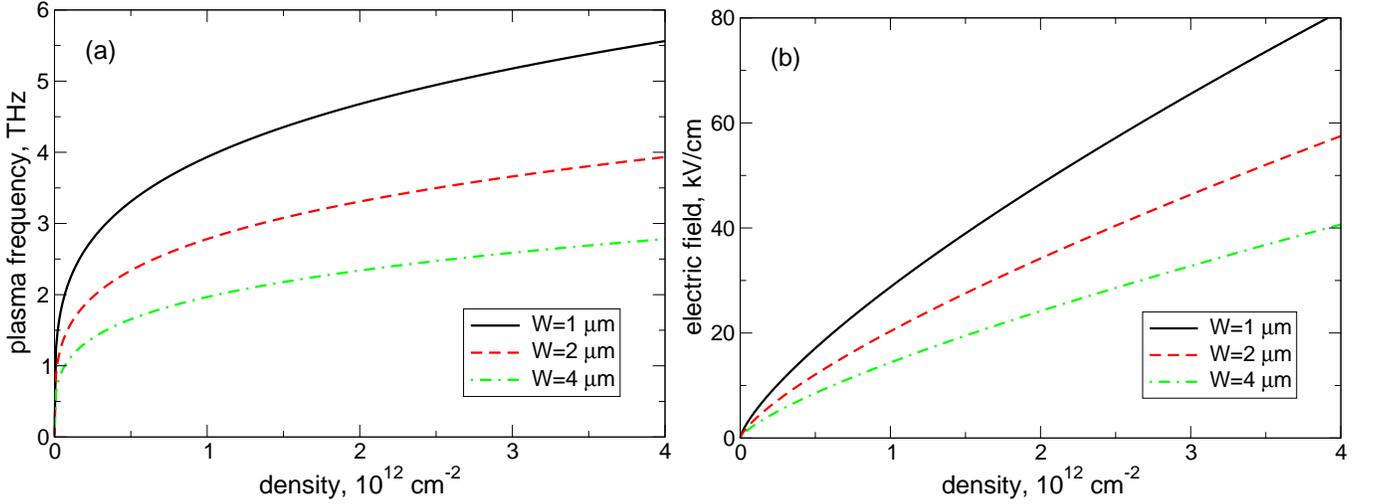

\includegraphics[width=0.495\textwidth]{Wpl.eps}
\includegraphics[width=0.495\textwidth]{NLfield.eps}
\caption{\label{fig:NLfield} (a) The 2D plasmon frequency determined by Eq. (\ref{plasm}) and (b) the nonlinear electric field determined by the condition ${\cal F}_p= 1$ as a function of the electron density $n_s$ at several values of the stripe width $W$. It is assumed that $q=\pi/W$ and $\kappa_0=4$.}
\end{figure}

Figure \ref{fig:NLfield} illustrates the density dependence of the plasma frequency (\ref{plasm}) and of the ``nonlinear'' electric field determined by the condition ${\cal F}_p= 1$ (i.e. the field required to observe the nonlinear effects), for parameters of Ref. \cite{Ju11}. One sees that dependent on the electron density $n_s$ and the stripe width $W$ the plasma frequency lies in the range from $\simeq 1$ to a few THz and the electric field at which the nonlinear effects can be observed is of the order of a few to tens kV/cm.

\subsection{Nonlinearity in a SNOM experiment}

\begin{figure}
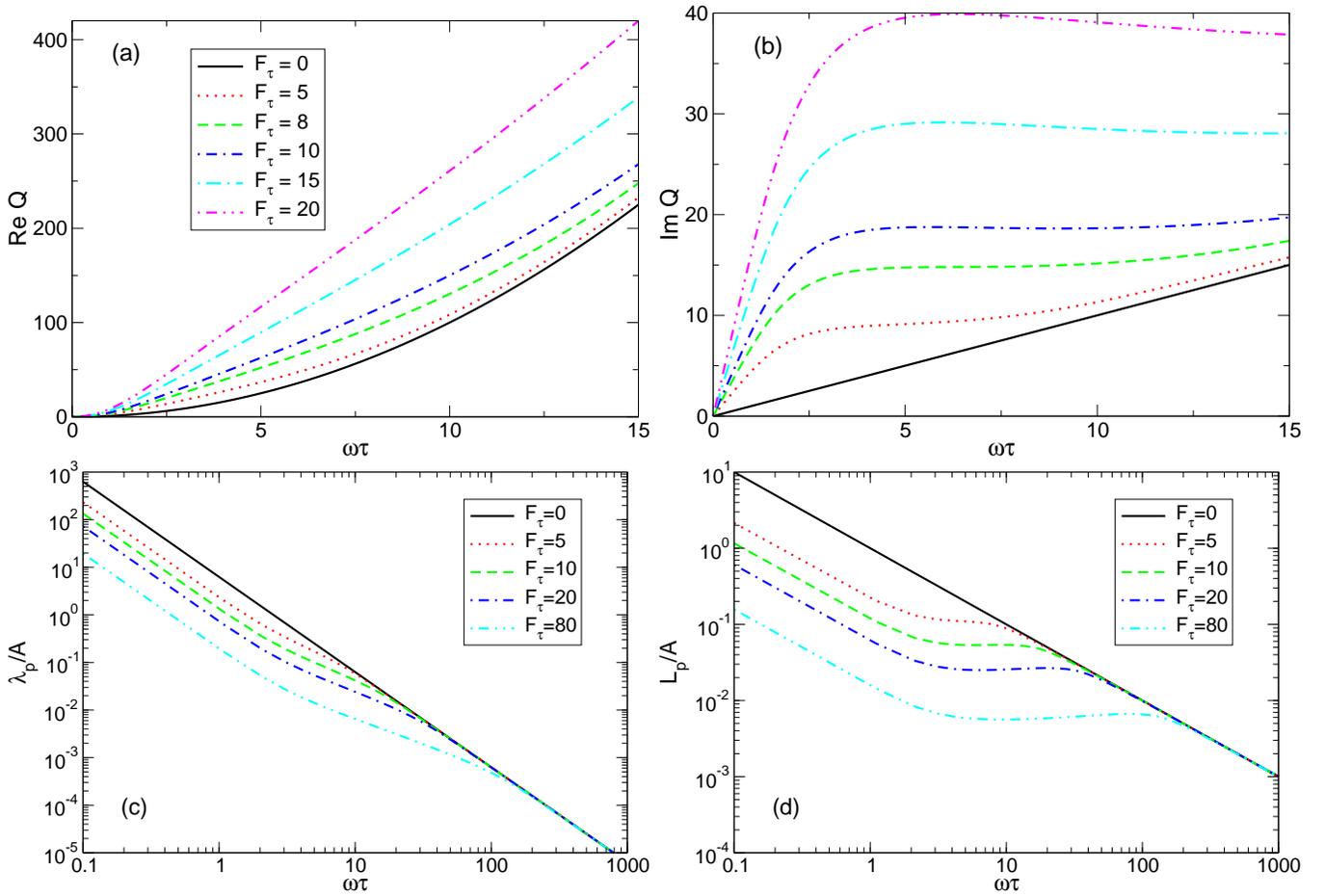

\includegraphics[width=0.495\textwidth]{ReQvsWt.eps}
\includegraphics[width=0.495\textwidth]{ImQvsWt.eps}
\includegraphics[width=0.495\textwidth]{LamAvsWt.eps}
\includegraphics[width=0.495\textwidth]{LpAvsWt.eps}
\caption{\label{fig:QvsW} (a) The real and (b) imaginary parts of the dimensionless wave-vector $Q$, Eq. (\ref{eqQ}), as well as (c) the wavelength $\lambda_p/A$ and (d) the propagation length $L_p/A$ of 2D plasmons as a function of $\omega\tau$ at a few values of the field parameter ${\cal F}_\tau$. The legends in (a) also refer to (b).}
\end{figure}

In a SNOM experiment the frequency $\omega$ is a fixed real value and the wave-vector $q$ is a complex function. In order to analyze its dependence on the frequency and field parameters we write it in the form
\be 
\frac  q{k_F}=\frac{\hbar v_F\kappa_0}{2e^2}
\left(\frac{\hbar}{  E_F\tau}\right)^2 Q,\label{qk}
\ee
and plot the real and imaginary parts of the dimensionless wave-vector 
\be 
Q=\frac{i\omega\tau}{{\cal S}_{1}(\omega\tau,{\cal F}_\tau)},
\label{eqQ}
\ee
in Figure \ref{fig:QvsW}. At ${\cal F}_\tau\to 0$ the real and imaginary parts of $Q$ depend on $\omega\tau$ quadratically and linearly, respectively, $Q'=(\omega\tau)^2$, $Q''=\omega\tau$. When ${\cal F}_\tau$ grows both $Q'$ and $Q''$ increase, i.e., the wavelength and the propagation length become shorter in the nonlinear regime. At ${\cal F}_\tau > \omega\tau$ the real part of $Q$ grows linearly with $\omega\tau$ in a broad range $\omega\tau\gtrsim 1$, Fig. \ref{fig:QvsW}(a). The imaginary part of $Q$ first linearly grows with $\omega\tau$ and then saturates at approximately $Q''\simeq 2{\cal F}_\tau$, Fig. \ref{fig:QvsW}(b). The 2D plasmon wavelength $\lambda_p$ and the propagation length $L_p$ can be obtained from (\ref{qk}) and written as

\be 
\lambda_p=2\pi A
 \frac{\left|{\cal S}_{1}(\omega\tau,{\cal F}_\tau)\right|^2}{\omega\tau{\cal S}_{1}''(\omega\tau,{\cal F}_\tau)},
\ee
\be 
L_p=A
 \frac{\left|{\cal S}_{1}(\omega\tau,{\cal F}_\tau)\right|^2}{\omega\tau{\cal S}_{1}'(\omega\tau,{\cal F}_\tau)},
\ee
where 
\be 
A=\frac{2e^2E_F\tau^2}{\kappa_0\hbar^2}\approx 777.8\ \mu\textrm{m}\times \frac{\sqrt{n_s[10^{12}\textrm{ cm}^{-2}]}(\tau[\textrm{ps}])^2}{\kappa_0}
\label{A}
\ee
is a prefactor with the dimensionality of length. The frequency dependencies of $\lambda_p/A$ and $L_p/A$ are shown in Figure \ref{fig:QvsW}(c,d). The nonlinearity effect is seen, again, at $\omega\tau\lesssim {\cal F}_\tau$, i.e., at ${\cal F}_\omega\gtrsim 1$. Both $\lambda_p$ and $L_p$ decrease under the action of the strong electric field with the propagation length being affected stronger. It is interesting that in a certain interval of $\omega\tau$ ($1\lesssim\omega\tau\lesssim {\cal F}_\tau$) the length $L_p$ becomes almost frequency independent, Figure \ref{fig:QvsW}(d).  

The absolute values of $\lambda_p$ and $L_p$ can be estimated from Figure \ref{fig:QvsW}(c,d) and Eq. (\ref{A}). For example, if $n_s=10^{12}$ cm$^{-2}$, $\tau=1$ ps and $\kappa_0=3.9$, the length $A$ is about 200 $\mu$m. Then, if the plasmon frequency is 2 THz we get $\lambda_p=7.96$ $\mu$m and $L_p\approx 15.92$ $\mu$m at ${\cal F}_\tau\to 0$ ($L_p/\lambda_p=2$) and $\lambda_p=1.02$ $\mu$m and $L_p=1.13$ $\mu$m at ${\cal F}_\tau=80$ ($L_p/\lambda_p=1.1$). At higher frequencies both $\lambda_p$ and $L_p$ are smaller (e.g. at 10 THz and ${\cal F}_\tau\to 0$ the lengths are $\lambda_p=0.32$ $\mu$m and $L_p\approx 3.2$ $\mu$m, $L_p/\lambda_p=10$), but the influence of the strong electric field is similar: at ${\cal F}_\tau= 80$ $\lambda_p=0.176$ $\mu$m and $L_p=1.32$ $\mu$m ($L_p/\lambda_p=7.5$). 

\section{Summary\label{sec:summary}}

We have theoretically studied the influence of the nonlinear effects on the spectrum of 2D plasmons in graphene. In the FIR transmission experiments the strong external electric field is predicted to lead to a substantial red shift and to a broadening of the plasmon resonance. In the SNOM experiments the nonlinearity is shown to result in the reduction of both the wavelength and the propagation length of 2D plasmons as compared to the linear regime. The characteristic electric fields needed for observation of the nonlinear effects are determined by the condition ${\cal F}_\omega=eE_0/\hbar k_F\omega\gtrsim 1$ meaning that the nonlinearity in the 2D plasmon spectrum is more important at low frequencies and in samples with low charge carrier density. The absolute values of the electric field causing the nonlinear effects in the 2D plasmon spectrum lie in the range from a few to a few tens of kV/cm. 

\acknowledgments

The author thanks Nadja Savostianova for useful discussions. The work has received funding from the European Union's Horizon 2020 research and innovation programme GrapheneCore1 under Grant Agreement No. 696656.

\providecommand{\latin}[1]{#1}
\makeatletter
\providecommand{\doi}
  {\begingroup\let\do\@makeother\dospecials
  \catcode`\{=1 \catcode`\}=2\doi@aux}
\providecommand{\doi@aux}[1]{\endgroup\texttt{#1}}
\makeatother
\providecommand*\mcitethebibliography{\thebibliography}
\csname @ifundefined\endcsname{endmcitethebibliography}
  {\let\endmcitethebibliography\endthebibliography}{}


\end{document}